\documentclass[11pt]{article}
\usepackage{epsfig}
\usepackage{graphicx}
\usepackage{epsfig}

\begin{document}

\title{\bf  Structure and stability of the Si(105) surface}

\author{C.V.~Ciobanu$^{1}$, V.B.~Shenoy$^1$, C.Z.~Wang$^2$ and K.M.~Ho$^2$ \\
$^1$Division of Engineering, Brown University, Providence, RI
02912 \\ $^2$Ames Laboratory -- U.S. Department of Energy and
\\ Department of Physics, Iowa State University, Ames IA 50011 }

\date{\today}

\maketitle

\bigskip
\bigskip
\bigskip
\bigskip

\begin{abstract}

\narrower {\small Recent experimental studies have shown that
well-annealed, unstrained Si(105) surfaces appear disordered and
atomically rough when imaged using scanning tunnelling microscopy
(STM). We construct new models for the Si(105) surface that are
based on single- and double-height steps separated by Si(001)
terraces, and propose that the observed surface disorder of
Si(105) originates from the presence of several structural models
with different atomic-scale features but similar energies. This
degeneracy can be removed by applying compressive strains, a
result that is consistent with recent observations of the
structure of the Ge/Si(105) surface.
 }
\bigskip


\end{abstract}
\newpage

The self-organized growth of Ge/Si quantum dots has been
investigated extensively for more than a decade, driven by their
potential applications as optoelectronic devices and nanoscale
memories. In the early stages of growth, the quantum dots that
have pyramidal shapes bounded by \{105\} facets evolve from
stepped mounds without encountering any energetic barriers for
their nucleation \cite{sutter, trompross}. The absence of
nucleation barriers has been explained  by a competition between
the strain-dependent, negative step formation energy and repulsive
step-step interactions that have a weak dependence on strain
\cite{apl, jmps}. The atomic configuration of the (105) facets has
also been elucidated; a rebonded step model for the Ge(105)
surface under mismatch strain was found to play a crucial role for
the stability of this surface \cite{apl, jap-prl, jap-ss,
italy-prl}.

While the structure of Ge/Si(105) surface has been recently
elucidated, the Si(105) surface shows intriguing features that are
not well understood. Experimental work by Tomitori {\em et al.}
\cite{tomitori}, Fujikawa {\em et al.} \cite{jap-prl} and Zhao
{\em et al.} \cite{china-Si105} reveal that Si(105) is atomically
rough even after careful annealing, and its STM image does not
display two-dimensional periodicity. The analysis of Zhao and
coworkers \cite{china-Si105} suggests the presence of a structure
for the Si(105) surface with large (001) facets and double-height
steps. While a model for this surface with double-height steps was
presented in \cite{china-Si105}, a study of its stability has not
been attempted.

Motivated by the recent STM investigations \cite{jap-prl,
china-Si105}, we search for reconstructions of Si(105) based on
(001) terraces separated by single- and double-height steps
oriented along $\langle 100 \rangle$ directions. In addition to
the currently accepted models of Si(105) (shown, for example, in
Ref.~\cite{jap-prl}), we have found a few other possible
structures based on double-height steps. We investigate the
stability of these novel reconstructions, and propose that the
roughness of the Si(105) surface is due to the coexistence of
different structures with surface energies that are very close to
each other. Furthermore, we find that compressive strain
particularly favors a certain rebonded structure over all the
others, which explains the atomically smooth and periodic
structure of the Ge/Si(105) surface \cite{jap-prl, tomitori}.

\begin{figure}[p]
  \begin{center}
  \includegraphics[width=5.0in]{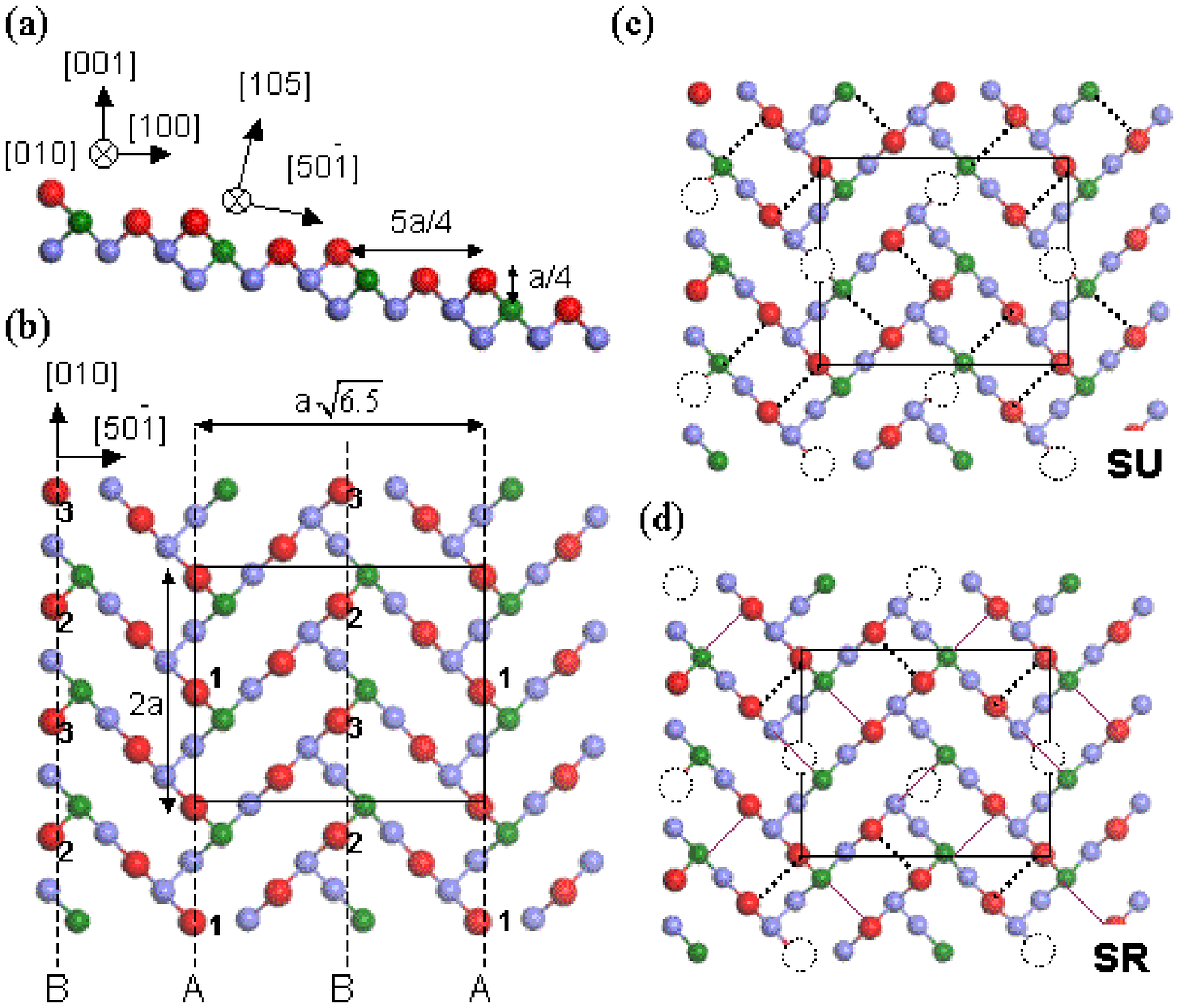}
  \end{center}
\vspace{-3cm}
\caption{Reconstructions of Si(105) with single-height steps and
(001) terraces. (a) Side view and (b) top view  of the
bulk-truncated (105) surface. Atoms are colored according to their
number of dangling bonds ($db$) before reconstruction: red=2$db$,
green=1$db$, and blue=0$db$. The (105) unit cells are marked by
rectangles in figs. (b)--(d). The single-height unrebonded (SU)
and single-height rebonded (SR) models are shown in (c) and (d);
these structures are obtained by eliminating atoms "1" and "2",
and atoms "1" and "3", respectively (refer to fig. (b)). The atoms
that are removed are shown as open circles in figs. (c)--(d). The
remaining atoms are bonded as indicated by black dotted lines
(dimer bonds) and purple solid lines (bridging bonds).}
\label{sh-reco105}
\end{figure}

To show how to obtain structural models for Si(105) in a
systematic manner, we start from the bulk-truncated structure and
attempt to reconstruct the surface in such a way that each atom
has at most one unsatisfied (dangling) bond after reconstruction.
We now present in detail the structures of Si(105) with single-
and double- height steps. The bulk-terminated Si(105), given in
Fig.~\ref{sh-reco105}(a), consists of Si(001) terraces of width
$5a/4$ separated by steps of monatomic height $a/4$, where $a$ is
the lattice constant of Si ($a=5.43$\AA). Terraces can be
reconstructed by forming short rows of dimers; because the dimer
rows are oriented at 45$^{\rm o}$ angles with respect to the
direction of the steps, every other atom on the step edges must be
eliminated in order to lower the number of dangling bonds ($db$).
Depending on the relative position of the step-edge atoms that are
eliminated, there are two distinct ways to achieve surface
reconstruction. With the notation adopted in
Fig.~\ref{sh-reco105}(b), one possibility is to eliminate atoms
"1" on edges A and atoms "2" on edges B, and to then form dimers
on the surface, as indicated by the dotted lines in
Fig.~\ref{sh-reco105}(c). This is the model originally proposed by
Mo {\em et al.} \cite{mo}, and later named PD (paired dimers)
\cite{jap-prl}. Because the atoms on the terraces do not rebond at
the step edges, we call this model SU (single-height, unrebonded)
\cite{apl}.

Another way to achieve single-height reconstruction is by
eliminating atoms "1" on edges A and atoms "3" on edges B
(Fig.~\ref{sh-reco105}(b)) and then by creating bonds between the
remaining surface atoms as indicated by the solid and dotted line
segments in Fig.~\ref{sh-reco105}(d). This model, which we call SR
(single-height rebonded)
%\footnote{Other groups used a similar
%name, RS (rebonded step) \cite{jap-prl, jap-ss, italy-prl}.}
%As a
%technical point, we note that the SU and SR models have the same
%number of atoms in the unit cell, which allows for a
% more accurate determination of their {\em relative}
%surface energy than used in \cite{jap-prl, jap-ss}.}
\cite{apl}, was proposed by Khor and Das Sarma \cite{kds} and has
recently been shown to appear on the side facets of the Ge quantum
dots \cite{jap-prl, jap-ss, italy-prl}. We note that there are two
different types of bonds on this surface: the usual dimer bonds
(dotted lines), and the {\em bridging} bonds (solid lines). The
bridging bonds join a two-coordinated atom and a three coordinated
one, leaving the former with only one dangling bond and fully
saturating the latter. Because the bridging bonds of the SR model
are stretched, this reconstruction is strongly stabilized by the
compressive strains present in Ge films deposited on Si surfaces
\cite{apl, jap-prl, jap-ss}.  Such rebonded models can also be
found for structures that have double-height steps, as discussed
below.

\begin{figure}[p]
  \begin{center}
    \includegraphics[width=5.0in]{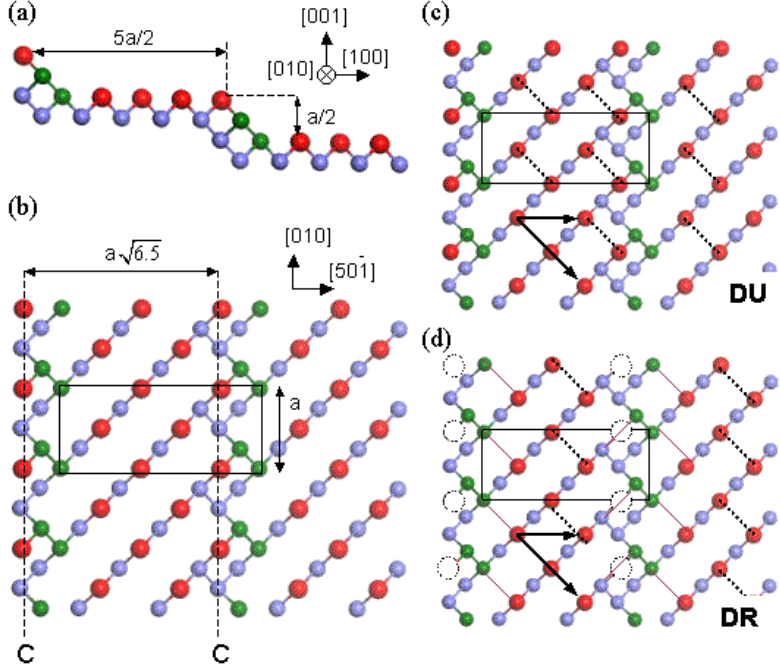}
  \end{center}
\vspace{-3cm}
\caption{Reconstructions of Si(105) with double-height steps and
$(\sqrt{2}\times1)$-Si(001) terraces. (a) Side view and (b) top
view of the bulk truncated (105) surface with double-height steps.
Atoms are colored according to the number of dangling bonds before
reconstruction as explained in Fig.~\ref{sh-reco105}. The (105)
unit cells are marked by rectangles in figs.~(b)--(d). The
double-height unrebonded (DU) and the rebonded (DR) structures are
shown in fig.~(c) and (d), respectively. The DR structure is
obtained after elimination of all the atoms at the step-edges
(rows C in (b), open circles in (d)), followed by dimerization
(black dotted lines) and rebonding (purple solid lines) . The
thick arrows in figs.~(c)--(d) represent the unit vectors of the
$(\sqrt{2}\times 1)$-reconstructed terraces.}
\label{dh-reco105-dudr}
\end{figure}

We now focus on the unreconstructed (105) surface with
double-height steps shown in Fig.~\ref{dh-reco105-dudr}(a)--(b).
Since the step height has doubled, the width of the terraces must
also be doubled, in order to preserve the overall surface
orientation. We found several models of double-height structures
of Si(105), with different terrace structures (e.g.
$(\sqrt{2}\times 1)$ or $(2\times 1)$ reconstruction) and atomic
bonding at the steps (e.g. rebonded or unrebonded), as explained
below.

The simplest model based on ($\sqrt{2}\times 1$) terraces is
illustrated in Fig.~\ref{dh-reco105-dudr}(c), where no atoms are
eliminated and dimers are formed as indicated by the dotted lines.
Since no rebonding is present, we call this the DU (double-height
unrebonded) structure. If we allow for the rebonding of atoms on
the lower terrace, then all the atoms on the step edges (denoted
by C in Fig.~\ref{dh-reco105-dudr}(b)) must be removed, so that
any surface atom would have fewer than two dangling bonds after
reconstruction. The remaining atoms are then bonded as indicated
by the solid and dotted lines in Fig.~\ref{dh-reco105-dudr}(d).
Like the SR model, there are two types of surface bonds, the
dimers and the bridges; we name this the DR (double-height
rebonded) structure.

\begin{figure}[p]
  \begin{center}
   \includegraphics[width=5.0in]{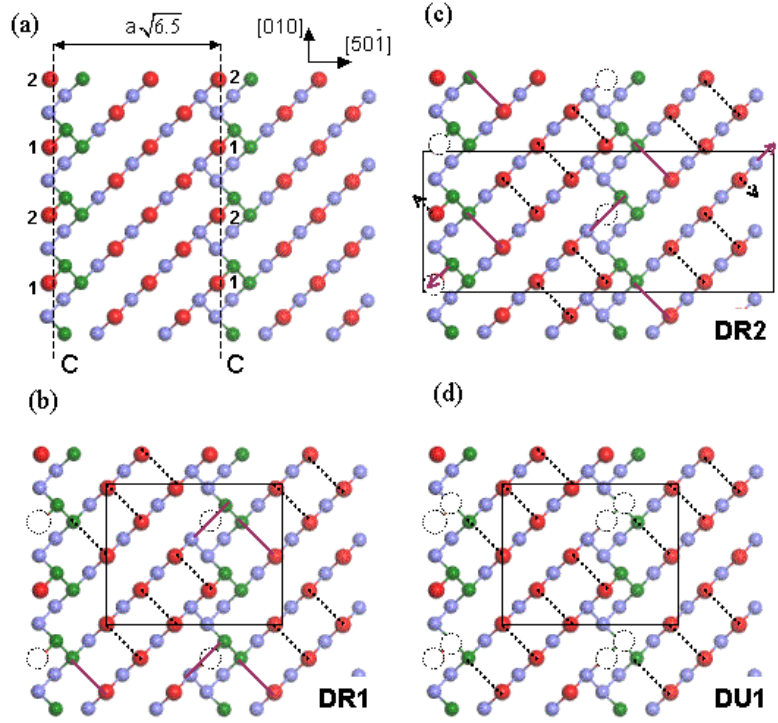}
  \end{center}
\vspace{-3cm}
\caption{Reconstructions of Si(105) with double-height steps and
$(2\times1)$-Si(001) terraces. Fig.~(a) shows the top view of the
bulk-truncated (105) surface with double-height steps. Atoms are
colored according to the number of dangling bonds before
reconstruction as explained in Fig.~\ref{sh-reco105}. Since the
dimer rows on the terraces are oriented at 45$^{\rm o}$ with
respect to the step edges C, every other atom on the step edges is
eliminated upon reconstruction. The elimination can proceed
in-phase (atoms "1" on all step edges) or out-of-phase (atoms "1"
on a given step edge and atoms "2" on the next edge). Since
rebonding at the step edges is present in both models (purple
solid lines), we label them DR1 (b) and DR2 (c). The atoms that
have been removed to achieve the reconstruction are shown as open
circles in figs.~(c)--(d). The (105) unit cells are marked by
rectangles with the dimensions $2a\times a\sqrt{6.5}$ for DR1 and
$2a\times 2a\sqrt{6.5}$ for DR2. The unrebonded model DU1
presented in Ref.~\cite{china-Si105} can be obtained by removing
one more atom from the DR1 unit cell as shown in fig.~(d). }
\label{dh-reco105-dr12}
\end{figure}

In the case of ($2\times 1$) reconstructed terraces, the dimer
rows are oriented at 45$^{\rm o}$ angles with the step edge. In
analogy to the single-height case, we can eliminate every other
atom on the step edges (rows C in Fig.~\ref{dh-reco105-dr12}(a)).
The elimination of atoms on consecutive terraces can be done
in-phase or out-of-phase, which leads to structures with different
periodic lengths in the [50$\overline{1}]$ direction, as shown in
Fig.~\ref{dh-reco105-dr12}(b) and (c). Since both of these models
involve rebonding at the step edges, we call them the DR1 and DR2
structures. From the DR1 structure we can obtain the unrebonded
model of Zhao {\em et al.} \cite{china-Si105} by removing another
atom from each unit cell, as illustrated in
Fig.~\ref{dh-reco105-dr12}(d); we label this the DU1 structure, to
distinguish it from the DU model in Fig.~\ref{dh-reco105-dudr}.

We have computed the surface energy of the structural models shown
in Figs.~\ref{sh-reco105}-\ref{dh-reco105-dr12} using two
empirical models for atomic interactions, namely the
Stillinger-Weber \cite{stillweb} and the Tersoff \cite{tersoff3}
potentials. While the empirical potentials provide a reasonable
description of stepped Si(001) surfaces (refer, for example, to
\cite{poon}), they are not able to capture the tilting (buckling)
of the dimers at the surface, which constitutes an important way
of surface relaxation for Si(105). Further, it is precisely the
tilting of the dimers that determines the major features of the
STM images and helps in the identification of the atomic structure
of the surface \cite{jap-prl, jap-ss, italy-prl}. In order to
capture the dimer tilting, we have used the charge self-consistent
tight-binding method of Wang {\em et al.} \cite{tbmd}, which
accurately predicts the energy ordering of several dimer-tilted
Si(001) structures \cite{tbmd111}. With this method, for each of
the Si(105) structural models described above, the total energy of
the atoms in the simulation cell exhibits many local minima, and
we search for the lowest energy structures by using a combination
of molecular dynamics simulations and annealing.

Our results are summarized in Table~\ref{gamma_at_zero_strain},
where the surface energies of the Si(105) models computed using
empirical \cite{stillweb, tersoff3} and tight-binding \cite{tbmd}
potentials are given. The table also contains the number of
dangling bonds per unit area for each model. We note that both the
Stillinger-Weber (SW) and the Tersoff (T3) potentials yield large
energy penalties for the dangling bonds on the surface, giving an
energy ordering similar to that predicted by bond counting. On the
other hand, the tight-binding (TB) description of atomic
interactions allows the structures with large numbers of $db$s per
area (SU, DU and DU1) to relax via dimer buckling, leading to an
entirely different ordering of the reconstructions. It can be seen
from Table~\ref{gamma_at_zero_strain} that the surface energies of
the (105) models are spread over an interval of 8--12 meV/\AA$^2$
when the empirical potentials are employed. In the case of
tight-binding, this interval is only $\approx$4meV/\AA$^2$, due to
a stronger relaxation of the unrebonded structures.

For all the potentials that have been used, we find that the SR
model has the lowest surface energy among {\em all} the
reconstructions considered; this is in agreement with previous
work \cite{apl,jap-prl, jap-ss, italy-prl} where only the SU and
SR models are compared. A closer look at the TB values in
Table~\ref{gamma_at_zero_strain} shows that the SR, DU and DR2
models have energies that fall within $\approx$1 meV/\AA$^2$ of
the energy of SU, indicating a near-degeneracy of the lowest
energy surface. Furthermore, our molecular dynamics simulations
show that there are other local minima (with different bond
bucklings) in the same energy interval. Due to this high
degeneracy, we propose that several models can be present
simultaneously on the Si(105) surface, which can explain the key
features observed in experiments -- disorder and roughness.
Because energy differences are small, the entropy associated with
the spatial distribution of different (105) unit cells would be
important even at low temperatures, explaining the lack of
2D-periodicity of the STM images \cite{jap-prl, china-Si105}; the
atomic-scale roughness observed in the STM images can be generated
by a random arrangement of the single- and double-height
structures that are close in energy. The proposal can be tested
experimentally by imaging (zooming in) different areas or
"patches" of the large-scale Si(105) samples. Preliminary work
along these lines has been reported in \cite{jap-prl,
china-Si105}.

\begin{figure}[p]
  \begin{center}
  \includegraphics[width=5.0in]{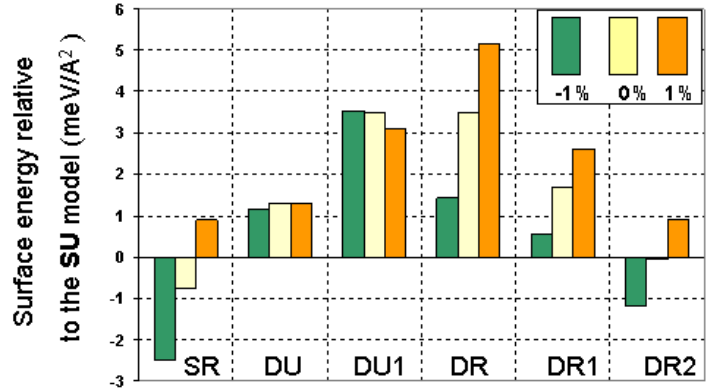}
  \end{center}
\vspace{-3cm}
\caption{Surface energies of different Si(105) structures measured
with respect to the surface energy of the SU model, computed with
the tight-binding method \cite{tbmd} for three values of applied
equibiaxial strain: +1\%(orange), 0\%(yellow) and -1\%(green).
While  there are three models (SR, DU, DR2) that have surface
energies within $\sim$1meV/\AA$^2$ of the surface energy of SU at
zero strain, a small amount of compressive strain (-1\%) removes
this near-degeneracy and strongly stabilizes the SR model over
{\em all} the other models.} \label{reco105-biax}
\end{figure}

We have also examined the strain dependence of the different
surface reconstructions. By calculating the energies of all the
model structures for three values of an applied equibiaxial strain
(-1\%, 0\% +1\%), we find that the near-degeneracy of the Si(105)
surface can be removed when the surface is subjected to a
compressive state of strain. As illustrated in
Fig.~\ref{reco105-biax}, even a compressive strain which is as
small as -1\% further stabilizes the SR model over the other
models. This result is supported by the recent observations of
Fujikawa {\em et al.}, who showed that the initially rough,
unstrained, Si(105) surface becomes smooth after being subjected
to a compressive mismatch strain through the deposition of 3
monolayers of Ge \cite{jap-prl}.

The origin of the strain-dependence of the surface energy lies in
the arrangement and type of the atomic bonds around the step
edges: if a structure contains significant rebonding (i.e. has
bridging bonds that are stretched compared to the bulk bonds),
compressive stresses tend to lower its surface energy. Indeed,
this trend emerges from the data presented in
Fig.~\ref{reco105-biax}: all rebonded structures show a decrease
of their energy (relative to SU) in compression. In contrast, the
energy gap between the unrebonded structures (DU and SU, DU1 and
SU) remains almost constant for the range of strains investigated
here. This finding is fully consistent with our previous work on
the formation energies of $\langle 100 \rangle$ steps \cite{apl},
as well as with the results of Refs.~\cite{jap-prl, jap-ss,
tomitori} on Ge/Si(105).

In summary, we have constructed a set of  structural models for
Si(105) and analyzed their stability using empirical
\cite{stillweb, tersoff3} and tight-binding \cite{tbmd}
potentials. Our study shows that the presence of single- and
double-height reconstructions on Si(105) can explain the
experimentally observed \cite{jap-prl, china-Si105} atomic
roughness and disorder of this surface. Three double-stepped
reconstructions (DU, DR1, DR2) were found to have lower surface
energies than the double-step model proposed in
\cite{china-Si105}. The atomic bonding at the step edge determines
the strain-dependence of the surface reconstructions and leads to
the strain-induced stabilization of the single-height rebonded
(SR) structure. Future experiments on strained Si(105) surfaces
(produced, for example, by bending) would be invaluable in gaining
further insight into the evolution of surface roughness as a
function of strain. The models presented here may also serve as
building blocks for other structures, for example, the quenched
($1\times 4$)-Si(105) observed in \cite{china-Si105}.

\bigskip{\bf Acknowledgements:} Support from the MRSEC at Brown
University (DMR-0079964), National Science Foundation grants
CMS-0093714 and CMS-0210095, and the Salomon Research Award from
the Graduate School at Brown University is gratefully
acknowledged. Ames Laboratory is operated by the U.S. Department
of Energy and by the Iowa State University under contract No.
W-7405-Eng-82. 

\newpage
\begin{table}[p]
\begin{center}
\begin{tabular}{l c c c c}
\hline \hline
\  &  SW           & T3         & TB        & Bond counting       \\
\  & meV/\AA$^2$ & meV/\AA$^2$ & meV/\AA$^2$  &$db/a^2\sqrt{6.5}$ \\
\hline
%
% 244-atom cell yields a TB value of 84.7534 meV/A^2 for the MO(SU) structure,
% and all others are computed with respect to this value (see excell file )
%
SU    & 99.63   & 99.40 & 83.54  & 6 \\
SR    & 90.39   & 90.79 & 82.78  & 4 \\
DU    & 102.24  & 99.36 & 84.84  & 6 \\
DU1   & 99.35   & 99.00 & 87.03  & 6 \\
DR    & 96.24   & 95.09 & 87.03  & 4 \\
DR1   & 96.27   & 96.64 & 85.22  & 5 \\
DR2   & 95.99   & 96.26 & 83.48  & 5 \\
%%%% The DU1 structure of Fig 6(b) in ref.\cite{china-Si105} has 6 $db/a^2\sqrt{6.5}$ %%%%
\hline \hline
\end{tabular}
\caption{Surface energy of Si(105) reconstructions calculated
using the Stillinger-Weber potential (SW) \cite{stillweb}, the
Tersoff potential (T3) \cite{tersoff3}, and the self-consistent
tight-binding  method  of Wang {\em et al.} (TB) \cite{tbmd}. The
last column indicates the number of dangling bonds ($db$) per
surface area expressed in units of $a^2\sqrt{6.5}$, where $a$ is
the bulk lattice constant of Si.} \label{gamma_at_zero_strain}
\end{center}
\end{table}

\end{document}